\input harvmac
\baselineskip=.55truecm
\Title{\vbox{\hbox{HUTP--96/A021}\hbox{hep-th/9606070}}}
{\vbox{\centerline{Anomalous Currents in 
SCFT$_4$}}}

\vskip .1in

\centerline{\sl A. Johansen\foot{E-mail: johansen@string.harvard.edu}}
\vskip .2in
\centerline{\it  Lyman Laboratory of Physics, }
\centerline{\it Harvard University, Cambridge, MA 02138, USA}
 \vskip .2in
\centerline{ABSTRACT}
\vskip .2in
   
We analyse\foot{Talk given at the Fourth International Conference on  
Supersymmetry SUSY96, May 29-June 1, University of Maryland, College Park.}
the critical behaviour of anomalous currents
in N=1 four-dimensional supersymmetric gauge theories
in the context of electric-magnetic duality.
We show that the anomalous dimension of the Konishi superfield 
is related to the slope of the beta function at the critical point.
We construct a duality map for the Konishi current in the minimal
SQCD.
As a byproduct we compute the slope of the beta function in the  
strong coupling regime. 
We note that
 the OPE of the stress tensor with itself does not close, but mixes 
with the Konishi operator.
As a result in superconformal theories in four dimensions (SCFT$_4$)
there are {\sl two} central charges;
they allow us to count both the vector multiplet and the matter multiplet 
effective degrees  of freedom.
Some applications to N=4 SYM are discussed.

\Date{\bf \phantom{December 1995}}

\newsec{Introduction}

A recent insight into quantum field theory has  been
the discovery by Seiberg of the electric-magnetic duality in non-trivial N=1 
superconformal
theories in four dimensions that can be realized as infrared fixed points of
ordinary N=1 supersymmetric theories \ref\emd{N. Seiberg, Nucl. Phys.
{\bf B 435} (1995) 129.}.

In this talk we discuss some aspects of critical behaviour of 
anomalous axial currents in the context of the 
electric-magnetic duality in N=1
supersymmetric theories\foot{This talk is based on the paper \ref\AGJ{D. Anselmi,  M. Grisaru and A. Johansen
{\it A Critical Behaviour of Anomalous Currents,
Electric-Magnetic Universality and CFT$_4$}.
Preprint HUTP--95/A048, BRX-TH-388; hep-th/9601023.}.}.

We demonstrate that  although the anomaly for the {\it renormalized} 
Konishi current in an off-critical theory is  
proportional to the
beta function, this anomalous current remains anomalous
at the conformal fixed point, and plays an important role in  
quantum superconformal
field theory in four dimensions (SCFT$_4$).
More specifically we show that the two-point correlator of  two
Konishi superfields behaves like $<J(x) J(y)>\sim
[\beta(g^2)/g^4]^2 |x-y|^{-4}$ where $g= g(|x-y|)$ is the running gauge coupling constant 
at the scale $1/|x-y|$ and
$\beta$ is the Gell-Mann-Low function.
Consequently in a theory with a non-trivial conformal fixed point
this correlator  has a power-like behaviour $\sim 1/|x-y|^{4+2\delta}$
at large distances, $|x-y|\to \infty$, where $\delta$ is given by
\eqn\res{
\delta =\beta' (\alpha_*) 
}
where $\alpha_*=g_*^2/4\pi$ is the critical value of the gauge coupling constant.
Eq. \res \ agrees with an analogous property that holds in two dimensions
\ref\cappelli{A. Cappelli and J. Latorre, Nucl. Phys. {\bf B 340} (1990)
659.}.
The components of the $J$ supermultiplet, in particular  the axial current $a_{\mu}$,
become conformal operators with  non-canonical conformal
dimensions.

We discuss some applications of this result to Seiberg 
electric-magnetic duality \emd .
We construct a duality map for the  Konishi operator.
Using the electric-magnetic duality \emd\ we argue that the
conformal dimension of the electric Konishi current is equal to that of its  
magnetic
counterpart in the dual formulation.
As a result we find a relation between the slopes of
the beta functions in these two formulations in the conformal window.

These features have their counterpart  in  the appearance of an additional
operator $\Sigma$ that mixes with the stress-energy tensor at the
level of OPE's and plays a fundamental role in SCFT$_4$. 
In the perturbation theory,
$\Sigma$ coincides with the Konishi current.
Consequently the OPE algebra in SCFT$_4$ is characterized by {\it two} central charges.
We discuss some aspects of OPE algebra in N=4 supersymmetric Yang-Mills theory.

The conserved currents have of course their canonical dimensions.
However, away from the fixed point,  the anomalous axial current that enters the 
supercurrent superfield
$J_{\alpha\dot{\alpha}}$ is actually a linear combination of
a conserved $R$ current and the Konishi current.
Therefore it contains  parts with different scaling behaviour.
As follows from the fact that the conformal dimension of the  
Konishi current
is larger then its canonical dimension, the Seiberg non-anomalous  
$R$ current
coincides with the current which enters $J_{\alpha\dot{\alpha}}$ in the  
infrared.
This last fact was first established in ref. \ref\ksv{I. Kogan, M. Shifman and  
A. Vainshtein, {\it Matching Conditions and
Duality in N=1 SUSY Gauge Theories in the Conformal Window}. Preprint
TPI-MINN-95/18-T, UMN-TH-1350-95, OUTP-95-25P, hep-th/9507170.}
by analysing operator equations for anomalous currents.

\newsec{Scaling dimension of  the Konishi current}

In order to determine the renormalization factor of an anomalous  
current it is convenient
to consider its matrix element in an external gauge field. 

For definiteness we shall consider  SUSY QCD with $SU(N_c)$  
gauge group and
 $N_f$ flavours of
 chiral superfields $Q_i$ and $\tilde{Q}^i$, $i=1,...,N_f$, in the
fundamental and antifundamental
representations, respectively.
One can define the Konishi
superfield $J=\sum_i ( \bar{Q}_i e^V Q_i + \tilde{Q}_i e^{-V} \bar{\tilde{Q}}_i) $
 which is singlet under the flavour global  
$SU(N_f)\times SU(N_f)$
group.
This current  is anomalous.

By differentiating the 1PI effective action $S_{eff}$
with respect to the bare coupling constant
$1/[g_0^2]$  (in the notation of ref. 
\ref\puz{M. Shifman and A.Vainshtein, Nucl. Phys. {\bf B 277} (1986) 456.})
in an appropriate kinematical regime \ref\nrjoh{A. Johansen,
Nucl. Phys. {\bf B 376} (1992) 432.}
\ref\MAnr{M.A. Shifman and A.I. Vainshtein, Nucl. Phys. {\bf B 365} 
(1990) 312.}
one gets 
\eqn\ww{
<{\rm Tr} \; W^2>= {\beta (\alpha)\over \beta_1  
(\alpha)}\cdot
{\beta_1
(\alpha_0)\over \beta (\alpha_0)}\cdot
{1\over 1-\alpha_0 N_c/2\pi} {\rm Tr} \; W^2_{ext} ,
}
where $\beta$ stands for total NSVZ beta function \ref\NSVZ{V. Novikov, M. Shifman, A.  
Vainshtein
and V. Zakharov, Nucl. Phys. {\bf B 229} (1983) 407;  {\it ibid}  
{\bf B 229}
(1983) 381;
Phys. Lett. {\bf B166} (1986) 329.},
$\beta_1$ is the one-loop beta function,
$\alpha=\alpha(k)=
g(k)^2/4\pi$ is the effective coupling at the scale of external
momenta, and $\alpha_0=g_0^2/4\pi$ is the bare coupling.
The factor $1/ (1-\alpha_0 N_c/2\pi)$ appears because of the one-loop form
of the Wilsonian action  \puz .

The anomaly in the Konishi current \ref\konishi{K. Konishi, Phys. Lett. {\bf B 135} (1984) 439.} 
is proportional to the operator ${\rm Tr} \; W^2$.
Hence, the renormalized matrix
elements of the Konishi current read 
\ref\AJ{A.A. Anselm and A.A. Johansen, Zh. Eksp. Teor. Fiz.  
Letters {\bf 49} (1989) 185; Zh. Eksp. Teor. Fiz. {\bf 96} (1989) 1181.}
\eqn\Jren{<\bar{D}^2 J_{ren} >={N_f\over 2\pi^2}\cdot
{\beta (\alpha)\over \beta_1 (\alpha)}W^2_{ext}.
}
This implies that the $z$ factor for the Konishi superfield
is proportional to $\beta (\alpha_0)/\beta_1 (\alpha_0)$.
To be precise, the total 
$z$ factor is  $z=(1-\alpha_0 N_c/2\pi) \beta (\alpha_0)/\beta_1 (\alpha_0) .$
Alternatively this fact can be deduced by differentiating the effective action for the chiral superfields with respect to a bare $Z_0$ factor in front of the kinetic terms $Q$ and $\tilde{Q}$ 
in the bare Lagrangian. 
Note that for N=1 abelian SUSY theories 
the invariance of $J(1-\gamma)$ 
(here $\gamma$ stands for the anomalous dimension of the fundamental chiral supermultiplets) under the renormalization group flow has been first established in ref. \puz .

Let us now consider the correlator of two Konishi currents in the
electric formulation of the theory.
By using
the Callan-Symanzik equation one can show that
\eqn\JJ{
<J(x) J(y)>\sim\left(\beta (\alpha)/
\beta_1(\alpha)\right)^2  ,
}
where $\alpha=g^2/4\pi$ is the coupling constant at the scale  
$1/|x-y|$. 

We consider the large distance limit of this correlator.
The factor $(\beta (\alpha)/\beta_1(\alpha))^2\to 0$ at $|x-y|\to \infty$
because $\alpha$ goes to the critical value $\alpha_*$.
Near the critical point the beta function is supposed to have a simple
zero, i.e. $\beta(\alpha)=\beta'(\alpha_*)(\alpha-\alpha_*).$
This means that the correlators are power-behaved at the critical point,
i.e.
\eqn\aa{
\alpha_*-\alpha=|x-y|^{-\beta'(\alpha_*)}\;\;{\rm at} \;\;  
|x-y|\to \infty .
}
Here $\alpha$ is taken at the scale $1/|x-y|$.
Substituting this expression into eq. \JJ \ we get
\eqn\JJJ{
<J(x) J(y)>\sim  
|x-y|^{-6-2\beta'(\alpha_*)} .
}
Let us now consider a non-asymptotically free theory with $N_f>3N_c$.
Assuming that such a theory flows into a free theory, which corresponds to 
$\alpha \to 0$, we see that the factor $\beta/\beta_1\to 1$ in the infrared.
The correlator of Konishi current at large distances has an integral  
power-like behaviour
which implies
that all currents have a canonical dimension as expected in a free  
theory.

\newsec{Konishi current in the magnetic theory}

We  consider  now the magnetic formulation
of the theory \emd , which has the $SU(N_f-N_c)$ gauge group, $N_f$
flavours of  dual quarks $(q^i,\tilde{q}_j)$
in the fundamental and anti-fundamental representations, respectively,
and
the meson field $M^i_j$ in the $(N_f,\bar{N_f})$ representation of the
flavour $SU(N_f)\times SU(N_f)$ group.
In contrast to the electric formulation this theory has a superpotential
$S=M^i_j q^i \tilde{q}_j$.
Thus the magnetic theory has two coupling constants.

We discuss the determination of  the magnetic counterpart of the Konishi current
of the electric theory.

It is convenient to consider an  extension of the electric theory which  
includes an additional
chiral superfield $X$ in the adjoint  representation of the gauge group
with a superpotential
$S_{el}={\rm Tr} X^3$.
Such a theory has been analysed in ref. \ref\kutasov{D. Kutasov,  
Phys. Lett. {\bf 351B} (1995) 230;
D. Kutasov and A. Schwimmer, Phys. Lett. {\bf 354B} (1995) 315;
D. Kutasov, N. Seiberg and A. Schwimmer, {\it Chiral Rings, Singularity Theory
and Electric-Magnetic Duality}. EFI-95-68, WIS/95/27, RU-95-75;
hep-th/9510222.}
and shown to flow to a non-trivial superconformal theory
(for $N_c/2< N_f < 2N_c$, see ref. \kutasov ).
On the magnetic side the theory includes the meson fields
corresponding to
$(M_j)^i_{\tilde{i}};\;\;\;
j=1,...,k .$
It also includes an additional
chiral superfield
$Y$ in adjoint representation of the dual gauge group $SU (kN_f-N_c)$.
The magnetic superpotential reads
$S_{mag}= \tilde{s} {\rm Tr} Y^{k+1} + \sum_{j=1}^k t_j  
M_j\tilde{q}Y^{k-j}q,$
where  $\tilde {s}$ and $t_j$ are coupling constants.

The strategy that we shall employ is to construct  non-anomalous Konishi currents
for the  dual versions of {\it this} theory, including the fields  $X$ and $Y$, and  
subsequently, by introducing  large mass terms for these
fields, recover the original  theory and corresponding currents in a low-energy limit.

In the electric theory we define a non-anomalous current which is a  
linear combination
of the Konishi current  and a current constructed from  the field $X$
(the conservation of this current is broken by the superpotential  $S_{el}$).
The corresponding non-anomalous current in the magnetic theory can be found by using 
holomorphicity of the Wilsonian action.
It turns out to be a particular linear combination of the currents of the fields
$q$, $\tilde{q}$, $M$ and $Y$.

In order to determine the duality map for the Konishi operator in the original SQCD 
we  add now a mass term  $m {\rm Tr} \;X^2$ and, via  
integration
over $X$ (or, equivalently, by the Appelquist-Carazzone  theorem)
go back to the  minimal  theory with fields $Q,$ $\tilde{Q}$  
on the electric side.
Also, by the duality map for chiral operators constructed in ref.  
\kutasov\  {\it at the critical point},    $Y$  will
have a mass term $m {\rm Tr} \;Y^2$  and we  will then reproduce the original  
theory in the
low-energy limit with the fields $q$, $\tilde{q}$ and $M$ on the  
magnetic side.
Moreover, to reproduce precisely the initial theory,
we have to choose a phase in which the gauge group is broken down to
$SU(N_f-N_c)$.
Note that the integration over the heavy field $X$ (and $Y$ on the  
magnetic side)
leads to a non-critical minimal SQCD because, in particular, the coupling constants
do not have  their critical values, e.g. $\alpha =\alpha_{\sharp}\neq \alpha_*$,
$\lambda=\lambda_{\sharp}$.
The effective Konishi operators in this effective non-critical SQCD  
will be obtained by dropping  out the heavy fields appearing in the non-anomalous  
Konishi operators
of the Kutasov model. 
The parameter $m$ is the only scale in the Kutasov theory and
plays the r\^ole of UV cutoff in this resulting non-critical
theory. 
The resulting Konishi operators should be thought of as bare operators
defined at this scale.
In order to describe their duality map 
in the critical SQCD we have to  analyse their  RG flow towards  
the infrared.

After dropping the fields $X$, $Y$, on the electric side the Konishi current $J_{el}$
has its usual form, and is expected to flow in the infrared to the corresponding
current in the critical minimal theory.  
In  the magnetic formulation an analysis shows that 
the magnetic counterpart of the Konishi operator  in the infrared
(of the minimal SQCD)
is a linear combination
$J_{mag}=A J_a +B J_{sp},$
where the currents $J_a$ and $J_{sp}$ are not conserved due to an anomaly and 
the superpotential respectively.
The values of $A$ and $B$ which are both non-zero, will  not be important for us.

Note that our identification of the magnetic counterpart for the  Konishi
supermultiplet implies that the operator ${\rm Tr} \;W^2$ in the  
electric theory matches on the magnetic side
with  a linear combination of the operator ${\rm Tr} \;W^2$ and the
superpotential, which is proportional to the superdivergence of the
operator $J_{mag}$.
Thus the duality map that we get for the anomaly multiplet
seems to be different from that of  ref. \ref\exam{See for a review
K. Intriligator and N. Seiberg, {\it Lectures on Supersymmetric Gauge
Theories and Electro-Magnetic Duality}. Preprint RU-95-48,  
IASSNS-HEP-95/70,
hep-th/9509066.} 
where it has been suggested
${\rm Tr} \;W^2_{el}=- {\rm Tr}\; W^2_{mag}$.

By analysing the matrix elements of the currents in the off-critical 
magnetic theory one can show that the $z$ matrix of renormalization of the 
non-conserved currents $J_a$ and $J_{sp}$ 
has the entries proportional to linear combinations of the beta 
functions of the two coupling constants of the model.
That means that the anomalous dimension of the magnetic Konishi current is given by the 
minimal eigenvalue $\beta_{min}'$
of the $2\times 2$ matrix of the slopes of the beta functions.
Combining this result with that for the electric theory we get
\eqn\bb{
(\beta' (\alpha_*))_{electric}= (\beta_{min}'( \alpha_*, \lambda_*))_{magnetic},
}
where $ \alpha_*$ and $\lambda_*$ are the critical values of the gauge and superpotential 
coupling constants, respectively.
It is interesting  to consider the strong coupling regime
$(2N_f-3N_c)/N_c<<1$.
In this case the magnetic theory is weakly coupled, we can compute $\beta_{min}'$
in perturbation theory, and thus we can explicitly
obtain the slope of the beta function in the strongly coupled  
electric theory
\eqn\bbb{
(\beta ' (\alpha_*))_{electric}={28\over 3}\left({3\over 2} -{N_f  \over N_c}\right)^2 .
}
In the next section we collect some simple observations about SCFT$_4$ that were 
stimulated by the investigation carried out so far. 
We shall see that  the Konishi current 
plays an important role in SCFT$_4$.

\newsec{OPE algebra}

We now study
SCFT$_4$ at the level of OPE's, and not simply at the level of
the classical conformal group.
It turns out that:

i) the OPE of the stress-energy tensor $T_{\mu\nu}$ with itself  
does not close \foot{OPE's in dimensions $2<d<4$ have been
studied recently by A. Petkou in ref. \ref\petkou{A. Petkou, Phys. Lett. {\bf B 359}
(1995)101 and hepth/9602054.}.};
another operator $\Sigma$ is brought into the algebra;

ii) there are {\sl two} central charges, $c$ and $c'$, one related to
$T_{\mu\nu}$, the other one to $\Sigma$; 

iii)  in general, $\Sigma$ has an anomalous dimension;  
it  coincides with the Konishi current $J$ in the perturbation theory.

More precisely in an N=1 superconformal theory we have 
\eqn\JJK{\eqalign{
J_{\alpha \dot{\alpha}}(z) J_{\beta \dot{\beta}}(z') =& c {X_{\alpha \dot{\alpha} \beta \dot{\beta}} \over (s^2 \bar{s}^2)^{3\over 2}}
+\Sigma (z') {Y_{\alpha \dot{\alpha} \beta \dot{\beta}} 
\over (s^2 \bar{s}^2)^{2-{h\over 2}}}
+ \cdots
,\cr
\Sigma (z) \Sigma(z')  =&{c'\over s^{2+h} \bar{s}^{2+h}} 
+\cdots ,
}}
where $J_{\alpha \dot{\alpha}}$ is the supercurrent,
$h$ stands for an anomalous dimension of the scalar (non-chiral) 
superfield $\Sigma$,
$X_{\alpha \dot{\alpha} \beta \dot{\beta}}$ and  $Y_{\alpha \dot{\alpha} \beta \dot{\beta}}$
are appropriate dimensionless tensor structures,
$
s_{\alpha \dot{\alpha}} = (x-x')_{\alpha \dot{\alpha}}  +{i \over 2}[ \theta_\alpha (\bar{\theta} - \bar{\theta}' )_{\dot{\alpha}} +
\bar{\theta}'_{\dot{\alpha} } (\theta - \theta ')_\alpha ] 
\bar{s}_{\alpha \dot{\alpha}} = (x-x')_{\alpha \dot{\alpha}}  +{i \over 2}[ \bar{\theta}_{\dot{\alpha} }({\theta} - {\theta}' )_{\alpha} +
{\theta}'_\alpha (\bar{\theta} -\bar{ \theta} ')_{\dot{\alpha} }]
$,
we use the conventions of
{\it Superspace}  \ref\Superspace{S.J. Gates, M.T. Grisaru, M. Ro\v{c}ek and W. Siegel,
{\it Superspace},  Benjamin-Cummings, Reading MA, 1983.}.

For an N=1 free theory with $m$ matter multiplets and $v$ vector  
multiplets, we have $c=15 v+ 5m$, $c'=5m$. 
For a generic interacting SCFT$_4$,   $c$ and $c'$ encode, via these formulae,
the effective numbers of  matter and vector multiplets, $m$ and $v$.

Consider now some applications to the N=4 supersymmetric Yang-Mills theory
which is conformal for any value of the gauge coupling constant $\alpha$.
Since $c$ is related to the conformal anomaly in an external  
gravitational field
\ref\CFL{A. Cappelli, D. Friedan and
J. Latorre, Nucl. Phys. {\bf B 352} (1991) 616.},
\ref\FB{See, for example, F. Bastianelli, Tests for C-theorems in 4D,
hep-th/9511065, and references therein.}
(which obviously does not have
any multiloop corrections in this theory)
it does not depend on $\alpha$.
In the weak coupling regime $c'=15n$, $n=v=m/3$.
It is not  clear whether $c'$ is also unaffected by turning on the interaction.
S-duality tells us that $c'=15n$ also in the infinitely strong  
coupling regime.
It would be interesting to identify $\Sigma$  in the interacting theory,
and compute the corresponding $c'$.  
Note that a deformation $\Sigma$
cannot preserve N=4  
supersymmetry, since
the only candidate in that case would be $\Sigma={\cal L}$, which is  
not the case in the free theory.
Thus, we expect $\Sigma$ to have  anomalous dimension  
$h\neq 0$.
Indeed in the weak coupling limit an explicit computation (using $\Sigma$ equal 
to the Konishi operator $J$) gives (for $SU(N_c)$ gauge 
group) $h=3\alpha N_c/\pi$.
In general due to $S$ duality the parameters $h$ and $c'$ must be 
real non-singular modular functions of 
$\tau=\theta/2\pi +4\pi i/g^2$;  $\theta$ is 
a theta angle in front of $F\tilde{F}$ in the Lagrangian (more precisely,
$\tau$ belongs to the fundamental domain of the $\Gamma_0 (N_c)$ subgroup of
SL$(2,{\bf Z})$ 
\ref\vw{C. Vafa and E. Witten, Nucl. Phys. {\bf B 431} (1994) 3.}).
A dependence of $c'$ on the coupling constant of N=4 theory would imply that
the effective numbers of chiral and vector multiplets (in the N=1 sense) changes 
with $g$ while the total number does not \ref\AFGJ{D. Anselmi, D. Freedman, M Grisaru
and A. Johansen, in progress.}.

The fact  that the parameter $h$ controls 
the scaling properties of
the chiral superfields suggests the following interpretation of the model.
The N=4 SYM theory may be viewed as a Wess-Zumino model coupled to an N=1 SYM.
One may be interested  
in the effective Wess-Zumino coupling constant 
``dressed'' by the gauge interactions. 
With such an interpretation 
$h$ plays a role of the ``dressed'' anomalous dimension 
of the Konishi current in the Wess-Zumino model which is related 
to the effective $\beta$-function as $h=\beta'-\beta/\lambda,$
where $\lambda$ stands for the effective superpotential coupling constant
\AFGJ .
This situation seems to be analogous to the gravitational dressing of 
the $\beta$-function in 2D gravity
 \ref\dr{See, for example, M. Grisaru and D. Zanon, Phys. Lett. 
{\bf B 353} (1995) 64, and references therein.}.

In conclusion, while the quantum conformal algebra  can provide us, 
through the identification of two SCFT$_4$'s with a map between
the two operators $\Sigma_{el}$ and $\Sigma_m$, the  
considerations of the
previous sections show how to map the Konishi currents. 
It is easy to see that $\Sigma$ is related to or even coincides with 
the Konishi operator (at least in the perturbation theory).
The  appearance of one
anomalous operator in the OPE of the stress-energy tensor helps to  
clarify the role of anomalous currents
in CFT$_4$.

{\bf Acknowledgements.}
The author is grateful to D. Anselmi and M Grisaru for collaboration,
and to D. Freedman for interesting and valuable discussions.
This work is partially supported by the
Packard Foundation and by NSF grant PHY-92-18167. 

\listrefs
\end